\documentclass[sigconf]{acmart} 

\usepackage{float} 
\usepackage{caption} 
\usepackage{quoting,xparse}
\definecolor{ForestGreen}{RGB}{34,139,34}
\usepackage{graphicx}
\usepackage{color}

\usepackage{hyperref}
\hypersetup{
    colorlinks,
    citecolor=blue,
    filecolor=blue,
    linkcolor=blue,
    urlcolor=blue
}

\NewDocumentCommand{\bywhom}{m}{
  {\nobreak\hfill\penalty50\hskip1em\null\nobreak
   \hfill\mbox{\normalfont(#1)}%
   \parfillskip=0pt \finalhyphendemerits=0 \par}%
}

\NewDocumentEnvironment{pquotation}{m}
  {\begin{quoting}[
     indentfirst=false,
     leftmargin=\parindent,
     rightmargin=\parindent]\itshape}
  {\bywhom{#1}\end{quoting}
  }
\AtBeginDocument{%
  }

\setcopyright{acmlicensed}
\copyrightyear{2025}
\acmYear{2025}
\acmDOI{XXXXXXX.XXXXXXX}

\acmConference[EASE 2025]{The 29th International Conference on Evaluation and Assessment in Software Engineering}{17–20 June, 2025}{Istanbul, Turkey}

\begin{document}

\title{Identifying Critical Dependencies in Large-Scale Continuous Software Engineering
}


\author{Anastasiia Tkalich}
\orcid{0000-0001-7391-4194}
\affiliation{%
  \institution{SINTEF}
  \city{Trondheim}
  \country{Norway}
}
\affiliation{%
  \institution{Software Engineering Research Lab (SERL) at Blekinge Institute of Technology}
  \city{Karlskrona}
  \country{Sweden}
}

\email{anastasiia.tkalich@sintef.no}

\author{Eriks Klotins}
\orcid{0000-0002-1987-2234}
\affiliation{%
  \institution{SERL at Blekinge Institute of Technology}
  \city{Karlskrona}
  \country{Sweden}
}

\author{Nils Brede Moe}
\orcid{0000-0003-2669-0778}
\affiliation{%
  \institution{SINTEF}
  \city{Trondheim}
  \country{Norway}
}
\affiliation{%
  \institution{SERL at Blekinge Institute of Technology}
  \city{Karlskrona}
  \country{Sweden}
}

\renewcommand{\shortauthors}{Tkalich et al.}

\begin{abstract}

Continuous Software Engineering (CSE) is widely adopted in the industry, integrating practices such as Continuous Integration and Continuous Deployment (CI/CD). Beyond technical aspects, CSE also encompasses business activities like continuous planning, budgeting, and operational processes. Coordinating these activities in large-scale product development involves multiple stakeholders, increasing complexity. 
This study aims to address this complexity by identifying and analyzing critical dependencies in large-scale CSE. 
Based on 17 semi-structured interviews conducted at two Nordic fintech companies, our preliminary findings indicate that dependencies between software teams and support functions, as well as between software teams and external entities, are the primary sources of delays and bottlenecks.
As a next step, we plan to further refine our understanding of critical dependencies in large-scale CSE and explore coordination mechanisms that can better support software development teams in managing these challenges.

\end{abstract}





\maketitle

\section{Introduction and Related Work}

Continuous Software Engineering (CSE) is becoming widely adopted in large software companies. The benefits of automation and faster feedback loops enable organizations to ship software faster and address customer needs more precisely~\cite{klotins2022continuous}. The adoption requires new tools and practices such as building automation and continuous integration and delivery, among others, to remove bottlenecks and ensure a steady stream of software from developers to end users. In addition, CSE includes non-development-related activities such as continuous planning, budgeting, and operational aspects like continuous use \cite{fitzgerald2017continuous}. These activities may occur at different organizational levels or outside the organization altogether; such as at the level of engineering, operations, product, and end users \cite{klotins2022towards}.

Coordinating high-cadence work between software developers and stakeholders at different levels in large companies requires efficient and effective coordination. Moreover, the extent of coordination needed may be such that coordination itself becomes a bottleneck. 
Traditional coordinating mechanisms, such as meetings and development sprints, do not scale well when the software delivery cadence increases, therefore bottlenecks can emerge in the interface between software and non-software parts of the organziation~\cite{forsgren2018accelerate, dikert2016challenges}. Organizations relying on CSE have a need for increased coordination comparing to organizations that follow other software delivery methods~\cite{norbjerg2024never}. 

At the same time, excessive coordination is considered wasteful and contributes to organizational conflicts~\cite{alahyari2019exploratory,klotins2023organizational}. The coordination overhead could hinder engineering productivity~\cite{cataldo2008socio}. In turn, inefficiencies and hindrances in the engineering process may lead to unnecessary context switching, reduced focus, and general dissatisfaction among engineers~\cite{forsgren2018accelerate}.

Agile frameworks such as SAFE~\cite{putta2018benefits} and LESS~\cite{girma2019agile} attempt to address some stakeholder coordination issues. Yet, the proposed solutions require careful planning and synchronization between software teams. Importantly, none of these frameworks address the fundamental challenges of coordinating work among different organizational levels, for example, between software engineering and the support functions, such as business planning, legal, marketing, etc. These levels may differ in their working methods, cadence, goals, and even language~\cite{dikert2016challenges}. 
Coordination across these levels often takes time, e.g., scheduling a meeting to resolve an issue creates excessive waiting time and thus goes against the spirit of CSE with high delivery pace~\cite{forsgren2017dora}.


Identification of critical dependencies in large-scale CSE is the first step towards a root-cause analysis of coordination bottlenecks, and devising better and improved methods for coordinating high-paced engineering work. In this paper, we present preliminary results from a qualitative survey with two large Nordic fintech companies. We use qualitative coding methods to analyze 17 interviews to answer our research question of: \textit{What are critical dependencies in large-scale CSE?}


\textbf{Coordination as managing dependencies.} Several theories conceptualize coordination in general and coordination in software engineering in particular~\cite{berntzen2022taxonomy}. One of the most cited perspectives defines coordination as managing dependencies to achieve common goals~\citep{okhuysen_coordination_2009,malone1994interdisciplinary}. That is, \textit{dependencies} introduce the need to coordinate when actors need to relate to each other to solve a task that concerns them all. According to Malone and Crowston ~\cite{malone1994interdisciplinary}, a dependency emerges when the progress of one action relies upon a timely output of a previous action or a specific thing, e.g., an artifact, a person, or a piece of information. For example, a developer waiting for a code review depends on her colleagues to be able to complete her work.

Based on the work of Malone and Crowston, Strode~\cite{strode2016dependency} proposed a taxonomy of dependencies in engineering projects and identifies process, resource, or knowledge dependencies with further sub-categories. A resource dependency could be either of a technical nature or an entity. Process dependencies are further categorized into activity and business process dependencies. Knowledge dependency comprises requirement, expertise, historical, and task allocation dependencies. 

\textbf{Coordination in large-scale CSE.} Building software in large organizations requires substantial coordination between various specialized teams \cite{edison2021comparing,berntzen2022taxonomy,berntzen2023responding}. The problem of inter-team coordination has been earlier addressed by the research on large-scale agile. A systematic literature review identified several related challenges, such as managing dependencies between the teams, balancing team autonomy with inter-team dependencies, challenges related to globally distributed work, and achieving technical consistency across the teams \cite{dikert2016challenges}. A study of 70 teams showed that requirement dependencies in large-scale agile affect software releases and increase work coordination complexity \cite{biesialska2021mining}.

Although inter-team coordination is difficult in agile, coordinating CSE on scale may be even harder. The extremely high cadence of software delivery in CSE introduces the need for even more efficient coordination between the dependent teams. If a team releases software several times a day, all the features that wait in line due to other teams and functions become lost potential. As a recent case study shows that teams transitioning to CSE may face additional dependencies related to new test and deployment processes, code reviews, and infrastructure \cite{norbjerg2024never}. Coordination in the context of DevOps happens through four categories of boundary objects, such as requirements, deployed system information, process checklists, and flexible-format artifacts \cite{matthies2023investigating}.   
 
Furthermore, large-scale CSE 
 touches upon non-software layers of the organization, such as the business development, the end-user and the strategic planning \cite{klotins2022towards,soares_effects_2022}. This complexity introduces the need for frequent stakeholder negotiations. Negotiation between stakeholders from different levels is not a trivial task and sometimes requires dedicated roles to do the job \cite{tkalich2022toward}. Such negotiations can become bottlenecks and lead to organizational conflicts. As shown by a study of three software organizations, such conflicts are likely to arise in companies with functional organizational structures, inimical, die-hard processes, a culture of sign-offs, and a lack of autonomy and support for conflict resolution \cite{klotins2023organizational}. 
    
Stakeholder negotiations 
have also been described as challenging in companies applying agile methods. Reasons include the inability of non-development functions to adjust to the incremental delivery pace, 
bureaucratic processes, and recurring silos \cite{dikert2016challenges}. 
With an even more rapid delivery pace, the problem of slow stakeholder negotiations becomes even more pressing. 

\section{Method}

To identify the critical dependencies in large-scale CSE we conduct a qualitative survey relying on data from two large-scale Nordic fintech companies, SpareBank 1 Utvikling and Gjensidige. Both companies have been focusing on the introduction of agile practices and methods for more than 10 years, promoting continuous improvement in software teams, and embraced hybrid software development \cite{conboy2023future}.

\textbf{SpareBank 1 Utvikling (SB1U)} is a leading Norwegian fintech company developing digital banking solutions for a network of Norwegian banks. Over the years, SB1U has transformed from a company with high employee turnover to one of Norway’s most innovative and attractive tech companies \cite{moe2023attractive}. As a large-scale agile organization, SB1U has structured its teams into product areas to enhance collaboration, coordination, and delivery speed. The bank has
60 software teams, each of which typically comprises five to six developers, a tester, a user experience designer, a product owner, and a team leader. 
These product areas focus on specific digital banking products, enabling better alignment between teams working on related solutions. 

\textbf{Gjensidige} 
is a leading Nordic insurance company, offering a range of insurance products, pensions, and savings. Before 2023, Gjensidige is known for having a mature agile environment in their software development department, but in 2024, the company underwent a major transformation, using terminology from the Spotify model as an inspiration to scale agility across the entire organization. The transformation was designed to improve alignment across business and technology, and streamline decision-making. As part of this shift, Gjensidige restructured its teams into tribes, squads (typically 8-12 team members), chapters, and guilds to enhance cross-functional collaboration, innovation, and speed in delivering customer-centric solutions. 
Additionally, the company continues to invest in automation, digitization, and AI-driven services to improve operational efficiency and customer experience.

\begin{table*}[ht]
  \small 
  \renewcommand{\arraystretch}{0.85} 
  \setlength{\tabcolsep}{2pt} 
  \caption{Cases and data collected}
  \label{tab:cases}
  \begin{tabular}{p{1.6cm} p{1.2cm} p{2.7cm} p{6cm}} 
    \toprule
    Company & Team ID & Role (n. of interviews) & Release frequency in the team (approximately) \\
    \midrule
    SB1U & A1 & TL (1), PO (1) & 4-8 times/day\\
    & A2 & TL (2) & 5-6 times/day \\
    & A3 & TL (2) & 4-5 times/day \\
    & A4 & TL (1) & Several times/day\\
    Gjensidige & B1 & TL (1), PO (1) & \\
    & B2 & TL (1), PO (1) & 2-4 times/week or whenever ready\\
    & B3 & TL (1), PO (1) & 5-6 times/day\\
    & B5 & TL (1), PO (1) & Bi-weekly or whenever ready\\
    & B6 & TL (1), PO (1) & 5-6 times/day or whenever ready\\
    \midrule
    Total & 9 teams & 17 interviews & \\
    \bottomrule
  \end{tabular}
\end{table*}


\subsection{Data sources} 
We conducted 17 semi-structured interviews from 9 teams ( Table~\ref{tab:cases}). The primary data collections took place in February-March 2025. We relied on purposive sampling, meaning that we recruited only the informants who worked in the context of CSE. Our primary targets were members of software teams because they are likely to have a firsthand understanding of where the bottlenecks occur. For the teams A1-A3, pre-interviews were conducted with the same participants (Jan, 24) to understand the team´s domain and routines. 

All teams interviewed in this study are software teams, meaning that they own a specific fragment of a software product that faces the customers, and have relative authority in changing this product. These teams operate as autonomous cross-functional teams and can make and deploy changes end-to-end independently of other units. However, in practice, they often depend on other teams and support functions to understand and validate the requirements and acquire access to resources, such as data and specific expertise. In this paper we refer to them shortly as \textit{software teams}.

The interview guide was structured to capture the majority of the dependencies taking place in the teams and between the teams and the external context, and to identify which of these dependencies were critical. We relied on the taxonomy of dependencies from Strode \cite{strode2016dependency} to formulate questions on each type of dependency. To prompt the participants to retrieve the critical dependencies, we asked questions such as “Where in your team do you experience the most delays?” and “What are the main obstacles to your progress?”. For transparency, our interview guide is available online\footnote{The full interview guide is available at: \url{https://doi.org/10.5281/zenodo.15034833}.}

\subsection{Data analysis}
We analyze notes and transcripts from the interviews using some of the qualitative coding techniques described by Saldaña \cite{saldana2021coding}. We apply structural coding on the notes from the semi-structural interviews, about 2 pages per interview, the total 34 pages of notes. Structural coding involves reading through the data and identifying the fragments that relate to the research question \cite{saldana2021coding}. Guided by the interview notes, we applied axial and pattern coding on the transcripts, which means that we were trying to make sense of the fragments and group them in a meaningful way. At this stage, we categorized the dependencies based on the level where they occurred (within teams, between teams, etc.) The codes were validated through a discussion between the co-authors and through cross-checking the codes with the actual interview transcripts.

\subsection{Threats to validity} \label{stenglim}
As the paper presents early-stage findings, several validity concerns should be acknowledged. First, it focuses only on a subset of “critical” dependencies rather than providing a comprehensive view, to better highlight those linked to delays and associated challenges. External validity is limited due to the small number of cases and interview roles; however, the issues identified are not unique to the Nordics or fintech, but reflective of broader industrial software engineering practices. Lastly, the analysis is primarily based on interview notes to expedite early results and refine the full-transcript analysis approach. Accuracy was supported through cross-checks with transcripts and involvement of multiple authors in both interviews and analysis, enhancing reliability.

\section{Results}

\begin{table*}[h]
    \small 
    \renewcommand{\arraystretch}{0.85} 
    \setlength{\tabcolsep}{2pt} 
  \caption{Critical dependencies reported in the case teams.}
  \label{tab:dependencies}
  \begin{tabular}{p{2.5 cm} p{13.5 cm} p{1.5 cm}} 
    \toprule
    Where does the critical dependency occur & Description of the dependencies & Reported in teams \\
    \midrule
    Within the software teams & \textbf{{D1. Code reviews in teams:}} the team must approve code changes made by other team members, leading to delays in people’s own tasks and context switching. The expertise needed for approval is not always on the team. & B1, B5 \\
    Between the software teams & \textbf{D2. Related products:} the code changes made by one team have implications for the product owned by another team. 
    & A1, B3, B6 \\
     & \textbf{D3. Multiple incoming requests:} Many incoming requests from other teams causing the dependent teams to wait in line for completing their tasks.   & A3 \\
     \\
      & \textbf{D5. Code reviews across teams}: the team must approve code changes made by other teams, leading to delays in the team’s own tasks and context switching. & B3\\
      & \textbf{D6. The platform team}: Some code edits (especially creating new features) have to be ordered from the platform team, creating unnecessary waiting. & B5 \\
      & \textbf{D7. The security team:} the changes related to the security layer (e.g., API gateway) should be ordered from a dedicated security team. Since there is only one person handling such requests on this security team team, the process causes significant delays. & B5\\
      & \textbf{D8. The AI-team:} new AI-funcitonality has to be built in collaboration with the AI-team & B5 \\
      & \textbf{D9. User Acceptance-expertise:} Person with specific user acceptance expertise is not part of the team and has to be borrowed from other teams to complete acceptance testing. & B5 \\
      Between the software teams and the supporting functions & \textbf{D10. Database:} The database edits and data must be ordered and planned in advance with the database responsible, resulting in inflexibility and waiting. & A1, A4\\
      & \textbf{D11. Security department:} medium to big feature changes have to be risk assessed and revised by the dedicated department on the matter of compliance with regulations. The revision process takes a lot of time and is unpredictable causing the team to prioritize minor features and code improvements. & A1\\
      & \textbf{D12. Legal:} the legal team reviews and signs off all changes. Waiting time can be extensive and unpredictable. & A3, A4, B3\\
      & \textbf{D13. Sales:} all new features should be validated by the sales team, requiring booking their time in advance. This leads to inflexibility and waiting. & B3 \\
      & \textbf{D14. Internal core system:} Changes in the core system are the team needs to request and wait for any changes from a dedicated unit. Since many teams depend on the core system, the requests involve significant waiting.& B2, B6\\
      &\textbf{D15. Business development:} The team needs to interact with the business development unit to understand the logic behind the insurance products and/or to validate the new features. Coordination with the business unit happens through scheduled meetings which causes additonal waiting & A1, B3, B5, B6 \\
      Between the software teams and external entities & \textbf{D16. External core system:} Changes in the core system need to be requested from an external supplier. Since the supplier has infrequent release cycle, this causes waiting, from days to months. & A1, A2, A3, B1, B5\\
      & \textbf{D17. End-users:} The pool of the end users is limited because of the limited number of the corporate customers. This results in delays with respect to user testing. & B6\\
    \bottomrule
  \end{tabular}
\end{table*}
This section presents preliminary findings on the critical dependencies identified in the study (see Table \ref{tab:dependencies}).
\subsection{Dependencies within software teams}
Very few critical dependencies were reported  within the software teams. Most of the teams agreed that the tasks handled within the teams require a marginal amount of waiting. For example, a tech lead expressed: \begin{pquotation}{B3, TL} The tasks that are handled within our team tend to have rather good speed.
\end{pquotation}

Although people in all teams were relying on code reviews within the teams, very few considered such reviews to be bottlenecks. Only two teams reported issues with code reviews (D1) indicating a critical dependency. Team B5 reported a high number of reviews needing to be approved within the teams. In Team B1, the code reviews did not cause waiting time. However, the expertise necessary for reviewing such requests was limited in the team, resulting in poor quality of the reviews. 

\subsection{Dependencies between software teams}

Although several critical dependencies were reported in this category, not all of them were associated with long waiting. For example, although one team reported that code reviews from another team caused delays and context switching, many other participants did not agree. For example, one said:

\begin{pquotation}{B6, TL} It [the pull requests] goes very fast. It has something to do with us sitting close to each other, so we solve it like face-to-face. 
\end{pquotation}

Half of the dependencies in this category were reported by one team (B5)
. 
These dependencies related to other software teams (mostly enabling teams), such as the platform, the security, and the AI-teams (D6-D8). 

Other dependencies in this category were between the software teams who had to relate to each other due to the nature of the product (D2. Related products) 
Such dependencies could cause a large number of code reviews from other software teams (D5), which in the case of team B3 created delays on the team´s tasks and context switching. 

Team A3 reported an interesting approach to managing cross-team dependencies. They received a constant flow of requests from other teams, causing significant context switching. To address this, they handled requests weekly and only when they aligned with their own weekly goals. While this strategy reduced context switching, it also risked creating bottlenecks for other teams.


\subsection{Dependencies between the software teams and the support functions}

This group of critical dependencies was associated with long waiting and lack of flexibility. Many of these dependencies were reported not only in several teams but also in both companies, indicating that these issues may be more common than those in the previous category. For example, the need to validate the changes with legal teams (D12) 
The tech lead from team A3 described  the interface with the legal team (e.g. to approve the removal of the existing features) as very time-consuming. He said:  

\begin{pquotation}{A3, TL} If there is one concrete thing that requires lots of waiting, it is when we need to involve people from a higher level, for example, legal.
\end{pquotation}

Another critical dependency was between software teams and business development units (D15), which was reported in four teams. Relating to the business development units was crucial for the teams to understand their products and their dynamics. However, such units were described as major bottlenecks because the coordination was managed through meetings that had to be planned in advance. A team lead explained that coordination with the business unit happened through scheduled meetings: 

\begin{pquotation}{B6, TL} What takes the most time is to figure out the business rules [...]. Then you need to set up meetings that can easily happen in only one week or two.
\end{pquotation}

Other dependencies that are worth commenting on relate to shared technical infrastructure, such as databases (D10) and core systems (D14). Units managing this infrastructure were described as slow and bureaucratic
For example, participants from team A1 had to order database edits from a dedicated unit because they did not have rights to do these changes themselves. The tech lead said:

\begin{pquotation}{A1, TL} It took 1,5 months of quarreling just to edit one row in the database.
\end{pquotation}

Two teams from Gjensidige described their dependency on the internal core system, a fundamental part of the architecture that underpins all other elements. Changes to this system require specialized expertise and involve high risk, so they are typically handled by dedicated experts or units. As many organizational entities rely on the core system, all change requests had to queue:

\begin{pquotation}{B2, PO} When we need to carry out development that has an impact and must be coordinated with our core system […]
It is a big challenge, and it takes a very long time, and it has to be prioritized against other people’s tasks.
\end{pquotation}

\subsection{Dependencies with external entities}

Some of the reported dependencies crossed the organizational boundaries. A dependency that created severe bottlenecks in both companies was toward the core system, which was operated by external suppliers (D16). Comparing to the internal core systems that were congested with various requests, the core systems managed by the suppliers were even slower. The infrequent release cycles and plan-driven processes caused teams to wait for the requested changes for weeks and even months, as reported by a tech lead:

\begin{pquotation}{B5, TL} ...They work very much like a waterfall process. [...]And they're planning in phases like.. so if we order something new, it could take anywhere between one week to three months before they deliver the functionality in the core system.
\end{pquotation}

The end-user was also described as a critical dependency by one team (D17). This team was developing a solution facing the corporate customers. Since the pool of such customer organizations is rather limited, finding end users in these organizations was difficult, leading to slow user testing. The tech lead explained: 

\begin{pquotation}{B6, TL} It is harder to find users to test things on, to run user interviews and such things [...] So the user insight takes quite some time.
\end{pquotation}

Overall, we can see that some dependencies are reported only by a single team (such as D3-D9, D1, D17), whereas others - by several teams. This may indicate that some bottlenecks are specific to the context of particular teams, while others are more common. At the same time, several dependencies were identified in both companies (D2, D12, D15, D16). All these dependencies, except one (D2) occur in the interface between the teams and the support functions or external entities. This suggests that such dependencies are likely to be more generic and may occur in other companies as well. 

\section{Discussion}


In this paper we investigate critical dependencies in large-scale CSE to identify where the CSE flow is most likely to disrupt. 


\textbf{What are the critical dependencies in CSE?} Our findings indicate that bottlenecks are less severe within the teams and more severe at the interface between the software teams and the support functions, and between the software teams and external entities, see Table~\ref{tab:dependencies}.

First, our results indicate that there are few bottlenecks within the software teams themselves, see D1, in Table~\ref{tab:dependencies}. The participants in our study agreed that the work within the teams is relatively seamless because the teams have developed a high level of teamwork, control over the technology, and expertise relevant to their products. Such a result is somewhat expected, since both case companies have been using agile for many years, promoting frequent releases and continuous improvement. Such changes have been mostly targeted at increasing development efficiency on a team level.

Further, we found that although several bottlenecks were reported in coordination across teams, such as waiting for the platform team to implement changes, they caused only moderate delays. The majority of these bottlenecks also seemed to be specific to particular teams (D3-9). 
The state-of-the-art identifies substantial challenges in coordinating work across multiple agile teams, such as defining team interfaces, balancing autonomy and alignment, synchronizing hybrid work, and achieving technical consistency~\cite{dikert2016challenges, vsmite2023decentralized}.

Although coordinating engineering work inside and especially across teams may be challenging, our results suggest that this challenge is far less critical than dependencies to support functions and. Our findings show that dependencies between software teams and support functions, such as legal, business development, and external entities, are the most critical. Due to different working methods, resolving such a dependency may delay work from several days to months. For instance, the process of coordinating changes in the central database or the core system were described as very slow and inflexible, see Table~\ref{tab:dependencies}, dependencies, see D10, D14 in Table~\ref{tab:dependencies}. 

Our results align with Dikert et al.~\cite{dikert2016challenges} identifying that the support functions often operate by different rules than the software teams and could impede the work of software teams in large-scale agile. For instance, the work cadence could be different (D12) and communicating through scheduled meetings (D15). The support functions often have insufficient throughput to handle spikes in incoming requests from development teams (D5, D14, D16).

Furthermore, monolithic legacy architectures have also been described by our participants as contributing to bottlenecks (D16). The core systems, such as the one reported in a case study by Mazzara et al. ~\cite{Mazzara20211464} at Danske Bank, are common to fintech and are mission-critical for financial operations. At the same time, as this case study shows, such systems tend to be complex and inflexible, thus reducing the agility of the software development process.  

Finally, our results show that some dependencies cross the organizational boundaries and meet external entities, such as end-users (D17) and sub-contractors (D16). One prominent example reported in both cases was the bottleneck that occurred when a part of the critical architecture (the external core system) was managed by an external provider, see D16. Such bottlenecks may be challenging to eliminate because they involve the entities beyond the teams' and even the organizational control. Parts of the banking core systems are sometimes owned by external providers, adding to the complexity of the legacy architectures \cite{Mazzara20211464}. Our findings highlight the importance of developing more optimal coordination mechanisms to manage dependencies with such external entities in CSE. 

\textbf{Practical implications.} Our preliminary findings suggest that large-scale organizations engaged in CSE should shift their improvement efforts beyond intra-team efficiency to enhance coordination with support functions and external entities. Specifically, bridging cultural and procedural gaps between development teams and support functions is essential, as differing cadences and communication practices often create critical bottlenecks. Additionally, addressing the constraints imposed by monolithic legacy systems through incremental modernization or architectural encapsulation can significantly improve responsiveness and agility across the software delivery pipeline.

\section{Conclusions and further work}
Studying dependencies in large-scale CSE companies may help identify where the software development process flow is likely to be disrupted. 
This study reports preliminary findings 
describing dependencies in two large-scale fintech companies. Our results indicate that the coordination challenges are less likely to occur within or between the development teams and more likely to occur in the interface between the teams and the support functions and between the teams and external entities. 

In the future, we aim to collect additional quantitative data, data from people holding other roles (designers, business, legal) and, potentially, other companies. This data will be combined with the data reported in this study and further analyzed and synthesized to categorize the dependencies and their root causes. Additionally, we are going map the dependencies to suitable coordination mechanisms and gauge their efficiency and effectiveness. This work will address some of the threats to validity described in Section \ref{stenglim} and also help understand the mechanisms behind the emergence of bottlenecks and explore coordination mechanisms that can support the CSE-teams in better managing the critical dependencies.  


\textbf{Acknowledgments}
 This study is funded by the Research Council of Norway through the project Vertigo (grant 346563), by Swedish Knowledge Foundation (KK-stiftelsen), and project SERT Profile (Ref. 2018/010). We thank our participants for sharing their valuable insights, and Gjensidige and SpareBank 1 Utvikling for their administrative support with data collection.   



\bibliographystyle{acm-reference-format}
\bibliography{./references.bib}

\end{document}